\documentclass[english,prl,twocolumn,superscriptaddress,amsmath,amssymb,a4paper]{revtex4}
\usepackage[T1]{fontenc}
\usepackage[latin1]{inputenc}
\usepackage{graphicx}
\usepackage[top=0.85in,bottom=0.9in,left=0.8in,right=0.75in]{geometry}
\usepackage{booktabs}
\usepackage{color}

\makeatletter
\usepackage{dcolumn}
\usepackage{bm}
\usepackage{epsfig}

\newcommand{\ignore}[1]{}

\usepackage{babel}
\makeatother

\usepackage{sidecap}
\sidecaptionvpos{figure}{c}

\begin{document}

\title{Kekul\'{e} Lattice in Graphdiyne: Coexistence of Phononic and Electronic Higher-Order Band Topology}

\author{Haimen Mu}
\affiliation{Hefei National Laboratory for Physical Sciences at the Microscale, CAS Key Laboratory of Strongly-Coupled Quantum Matter Physics, Department of Physics, University of Science and Technology of China, Hefei, Anhui 230026, China}

\author{Bing Liu}
\affiliation{Hefei National Laboratory for Physical Sciences at the Microscale, CAS Key Laboratory of Strongly-Coupled Quantum Matter Physics, Department of Physics, University of Science and Technology of China, Hefei, Anhui 230026, China}

\author{Tianyi Hu}
\affiliation{Hefei National Laboratory for Physical Sciences at the Microscale, CAS Key Laboratory of Strongly-Coupled Quantum Matter Physics, Department of Physics, University of Science and Technology of China, Hefei, Anhui 230026, China}

\author{Z. F. Wang}\thanks{E-mail: zfwang15@ustc.edu.cn}
\affiliation{Hefei National Laboratory for Physical Sciences at the Microscale, CAS Key Laboratory of Strongly-Coupled Quantum Matter Physics, Department of Physics, University of Science and Technology of China, Hefei, Anhui 230026, China}

%\pacs{73.43.-f, 73.22.-f, 72.25.-b, 71.20.-b}

%73.43.-f 	Quantum Hall effects
%73.22.-f 	Electronic structure of nanoscale materials and related systems
%72.25.-b   Spin polarized transport
%71.20.-b 	Electron density of states and band structure of crystalline solids

%\date{\today}% It is always \today, today, % but any date may be explicitly specified

\begin{abstract}
The topological physics has been extensively studied in different kinds of bosonic and fermionic systems,
ranging from artificial structures to natural materials. However, the coexistence of topological phonon and electron in one
single material is seldom reported. Recently, graphdiyne is proposed to be a two-dimensional (2D)
electronic second-order topological insulator (SOTI). In this work, based on density-functional tight-binding
calculations, we found that graphdiyne is equivalent to the Kekul\'{e} lattice, also realizing a 2D phononic
SOTI in both out-of-plane and in-plane modes. Depending on edge terminations, the characterized topological
corner states can be either inside or outside the bulk gap, which are tunable by local corner potential.
Most remarkably, a unique selectivity of space and symmetry is revealed in electron-phonon coupling
between the localized phononic and electronic topological corner states. Our results not only demonstrate
the phononic higher-order band topology in a real carbon material, but also provide an opportunity to
investigate the interplay between phononic and electronic higher-order topological states.
\end{abstract}

\maketitle
Over the past few years, the higher-order topological insulator (TI), as an overlooked intriguing
topological phase, has inspired great research interests \cite{1,2,3,4,5,6,7,8,9,10,TI1,TI2,TI3,TI4,TI5,TI6,TI7,TI8,TI9,TI10,TI11}. Different to conventional
or the first-order TI, the higher-order band topology of \textit{m}-dimensional \textit{n}th-order
TI is characterized by gapless states at (\textit{m-n})-dimensional boundary \cite{1}, greatly enriching the
fundamental knowledge of bulk-boundary correspondence in topological materials. Currently, there are
three different physical mechanisms that have been proposed to realize the SOTI. The first is breathing-lattice
mechanism, which can induce a nontrivial gap by lifting the band degeneracy between two folded Dirac bands \cite{11,12}.
This type of 2D SOTI has been experimentally confirmed in a variety of artificial lattices \cite{13,14},
including the acoustic \cite{15,16}, photonic \cite{17} and elastic \cite{18} systems.
The second is double-band-inversion mechanism, which can induce a nontrivial gap by inverting the bands with opposite
parity twice \cite{19}. This type of 3D SOTI has been experimentally confirmed in very few electronic materials,
including the Bi \cite{20}, Td-WTe$_2$ \cite{21} and Bi$_4$Br$_4$ \cite{22}. The third is in-plane Zeeman-field mechanism,
which can induce a nontrivial gap by breaking the helical Dirac edge or surface states of the first-order TI \cite{23,24,25,26,27}.
This type of SOTI is a magnetic SOTI that has not been experimentally confirmed yet. Nowadays,
it's emergent to search more SOTI candidates in natural materials for accelerating the experimental detection.

As a fast-booming field, the TI research in electron is also extended to
phonon in recent years \cite{28,29,30,31,32}. Since phonon is boson without spin-orbital coupling (SOC), the
well-known SOC mechanism for opening the nontrivial gap of electronic TI \cite{33} is invalid for
phononic TI. Due to such a discrepancy between phonon and electron, so far, there is no report for the
phononic and electronic TI in the same system. One notices that SOC is prerequisite for the first-order TI,
but not necessary for the higher-order TI \cite{13,14}. The breathing-lattice mechanism
provides a simple strategy to design electronic SOTI without SOC and phononic SOTI in the same framework,
making a chance to realize them simultaneously. Recently, Kekul\'{e} lattice, as one of the
breathing-lattices, is successfully created by intercalating Li atoms in graphene \cite{34}.
However, it is still a challenging task to realize an intrinsic Kekul\'{e} lattice without external modifications,
because this breathing order is not an energy-favorable configuration in most natural materials.
Additionally, from the symmetry perspective, the vibration basis of a phonon Hamiltonian can be viewed as
$p$ orbital \cite{35}, so the out-of-plane and in-plane phonon bands should be comparable to $p_z$ and $p_{x,y}$
electron bands in Kekul\'{e} lattice. This feature has already been observed in previous
model calculations \cite{12,30,35,36}, indicating the carbon-based Kekul\'{e} lattice to be an ideal
platform to study both phononic and electronic SOTI.

%\begin{SCfigure*}
\begin{figure*}
\centering
\includegraphics[width=16.8cm]{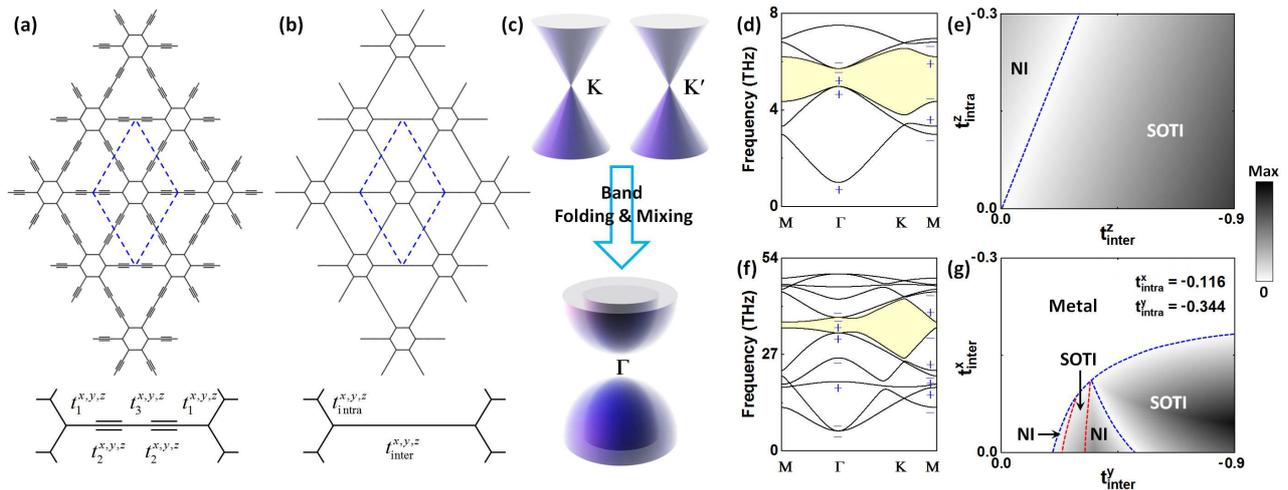}
\caption{(a) Atomic structure of graphdiyne and definition of out-of-plane ($t^z$) and
in-plane ($t^{x,y}$) nearest-neighbor force-constant. (b) Kekul\'{e} lattice in
simplified graphdiyne and definition of intra-cell ($t_{\rm{intra}}^{x,y,z}$)
and inter-cell ($t_{\rm{inter}}^{x,y,z}$) force-constant. The blue-dashed line is the
unit-cell in (a) and (b). (c) Schematic band folding and mixing induced Dirac gap at
$\Gamma$ point in the Kekul\'{e} lattice. (d) Out-of-plane phonon bands of Kekul\'{e} lattice with
$t_{\rm{inter}}^z=-0.006$, $t_{\rm{intra}}^z=-0.004$ and on-site $\varepsilon^z=0.014$. (e) Topological
phase diagram of out-of-plane phonon bands. The blue-dashed line is phase boundary,
where the band gap is closed. (f) In-plane phonon bands of Kekul\'{e} lattice with $t_{\rm{inter}}^x=-0.019$,
$t_{\rm{inter}}^y=-0.288$, $t_{\rm {intra}}^x=-0.116$, $t_{\rm {intra}}^y=-0.344$ and
on-site $\varepsilon^{x,y}=0.63$. (g) Topological phase diagram of in-plane phonon bands. The
blue-(red-)dashed line is phase boundary, where the band gap is (doesn't) closed.
The even/odd parity is labeled as $+/-$ and SOTI gap is highlighted by light-yellow color in (d) and (f).
The black- and white-color denotes the maximum and zero value of the gap in (e) and (g).
The unit of all force-constants is Hartree/Bohr$^2$ in (d)-(g).}
\end{figure*}
%\end{SCfigure*}

In this work, starting from the 2D graphdiyne, an electronic SOTI proposed recently \cite{37,38},
it is further identified to be a phononic SOTI by density-functional tight-binding calculations \cite{39},
demonstrating the coexistence of 2D phononic and electronic SOTI in one natural material.
Firstly, we show the equivalence between Kekul\'{e} lattice and graphdiyne, and map out a topological
phase diagram for both out-of-plane and in-plane modes, where an anomalous topological phase transition
without closing the bulk gap is found for the in-plane modes. Then, the edge termination dependent topological
edge and corner states are investigated systematically, where the robust topological corner states can be either inside
or outside the bulk gap, showing tunable energies by local corner potential. Lastly, the unique interplay
between phononic and electronic topological corner states is revealed by exploring the electron-phonon coupling (EPC),
where an enhanced EPC is achieved for phonon with even mirror symmetry coupled to electron
spatially localized at the same corner.

The atomic structure of 2D graphdiyne is shown in Fig. 1(a), which constitutes benzene
rings connected with alkyne unites \cite{40}. If we focus on the carbon atoms in benzene rings only, it is equivalent
to a Kekul\'{e} lattice, as shown in Fig. 1(b). Following the two-band analysis for
graphyne \cite{41}, the indirect phonon hopping between two vertex carbons through C-C single-bond (labeled as $t_1$
and $t_3$) and C$\equiv$C triple-bond (labeled as $t_2$) can be simplified to an effective inter-cell
hopping for both out-of-plane ($z$) and in-plane ($x,y$) force constant as $t_{\rm{inter}}=t_1^2t_3/t_2^2$ \cite{SI},
as shown in Fig. 1(a) and 1(b). From this kind of view, the phononic band topology of
graphdiyne will be captured by the Kekul\'{e} lattice, which is supported by our numerical calculations later.
Physically, the band gap of Kekul\'{e} lattice is induced by folding $\rm K$ and $\rm K^\prime$ Dirac
cone to $\Gamma$ point, and mixing them together through different inter-cell ($t_{\rm{inter}}^{x,y,z}$)
and intra-cell ($t_{\rm{intra}}^{x,y,z}$) hopping \cite{13,31}, as shown schematically in Fig. 1(c).
In principles, this gap opening mechanism is applicable to both out-of-plane and in-plane phonon bands.
However, the previous model of phononic and electronic SOTI are only limited in out-of-plane mode \cite{42}
and $p_z$ orbital \cite{12}, without considering the possibility of in-plane mode and $p_x,p_y$ orbital.

To fill out the missing region and fully understand the topological physics of SOTI in Kekul\'{e} lattice, the decoupled out-of-plane and in-plane
phonon bands are calculated separately, as shown in Fig. 1(d) and 1(f), where even/odd parity is labeled as $+/-$ at time-reversal-invariant-momenta (TRIM)
$\Gamma$ and M points. The quadrupole moment ($q_{12}$) is used as a topological invariant to distinguish the 2D SOTI \cite{12,43},
which is nontrivial (or trivial) for $q_{12}=1/2$ (or $0$) if the number of subbands with opposite parity between
$\Gamma$ and M below the bulk gap is 2,6,10 $\cdot\cdot\cdot$ (or 4,8,12 $\cdot\cdot\cdot$) \cite{SI}. Clearly, there are only two
subbands satisfy this rule below the light-yellow colored gap in Fig. 1(d) and 1(f), exhibiting a 2D SOTI phase
for both out-of-plane and in-plane modes. Furthermore, the topological phase diagram is mapped out to
show the parameter dependence of phononic SOTI. For the out-of-plane modes described by $t_{\rm{inter}}^{z}$
and $t_{\rm{intra}}^{z}$, the phase transition between SOTI and normal insulator (NI) is characterized by
closing the bulk gap at $t_{\rm inter}^z$=$t_{\rm intra}^z$ (blue-dashed line), as shown in Fig. 1(e).
The SOTI exists in space for $t_{\rm inter}^z$ larger than $t_{\rm intra}^z$, which is the
same to the electronic SOTI with $p_z$ orbital \cite{12}. However, the phase diagram is more complex for the
in-plane modes described by $t_{\rm{inter}}^{x,y}$ and $t_{\rm{intra}}^{x,y}$. Besides the metallic phase without global gap,
there are two different types of phase transition between SOTI and NI, companied with (blue-dashed line)
and without (red-dashed line) closing the bulk gap, as shown in Fig. 1(g). The appearance of such an
anomalous phase transition without closing the bulk gap can be attributed to the special definition of quadrupole moment.
Different to the $Z_2$ index that accounts for the number of subbands with odd parity at TRIM \cite{44},
the quadrupole moment only accounts for the number of subbands with opposite parity between different TRIM \cite{12}.
Hence, the sequence of subbands below the bulk gap is meaningless in $Z_2$ index but meaningful in quadrupole
moment, which can induce an anomalous higher-order topological phase transition without closing the bulk gap.

%\begin{SCfigure*}
\begin{figure}
\centering
\includegraphics[width=8cm]{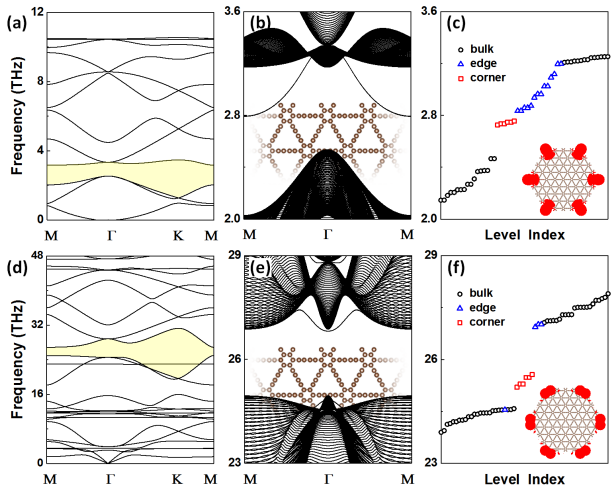}
\caption{(a) Out-of-plane phonon bands of graphdiyne with SOTI gap highlighted by light-yellow color.
(b) Out-of-plane phonon bands of graphdiyne nanoribbon with edge termination shown in the inset.
(c) Discrete out-of-plane phonon levels of hexagonal graphdiyne cluster with spatial distribution
of six corner states shown in the inset. The radii of red-circle denotes the intensity of phonon mode.
(d)-(f) are the same to (a)-(c), but for the in-plane modes of graphdiyne.}
\end{figure}
%\end{SCfigure*}

After establishing the universal phononic phase diagram of Kekul\'{e} lattice, its material
realization in 2D graphdiyne is studied directly. From the bulk aspect, the out-of-plane phonon bands of
graphdiyne are shown in Fig. 2(a), where the lowest six subbands are comparable to the Kekul\'{e}
lattice bands shown in Fig. 1(d). Focusing on the light-yellow colored gap, we found that its quadrupole
moment is $q_{12}=1/2$ \cite{SI}, showing a nontrivial bulk topology for 2D SOTI. From the edge aspect, the out-of-plane
phonon bands of graphdiyne nanoribbon are shown Fig. 2(b), where the edge is terminated by benzene ring connected
with two carbon atoms (inset of Fig. 2(b)). There is one edge state inside the bulk gap connected with
the upper bulk bands, showing the character of gapped topological edge state for 2D SOTI. From the corner aspect,
the discrete out-of-plane phonon levels of hexagonal graphdiyne cluster are shown in Fig. 2(c). There are six almost
degenerate corner states (red cube) inside the bulk gap, just below the discrete edge states (blue triangle).
The spatial distribution of them is localized at six corners (inset of Fig. 2(c)), showing the
character of in-gap topological corner state for 2D SOTI. Combining these three characters,
the 2D phononic SOTI is identified in the out-of-plane modes of graphdiyne.

Moreover, the higher-order topology for the in-plane modes of graphdiyne is also describable by
the Kekul\'{e} lattice. As shown in Fig. 2(d), the main feature of in-plane phonon bands of graphdiyne is comparable
to the twelve-band Kekul\'{e} lattice (Fig. 1(f)), where a quantized quadrupole moment ($q_{12}=1/2$) is obtained in
the light-yellow colored gap \cite{SI}. For the graphdiyne nanoribbon (inset of Fig. 2(e)), there are one pair of
gapped topological edge states inside the bulk gap, as shown in Fig. 2(e). For the hexagonal graphdiyne cluster,
there are six spatially localized in-gap topological corner states inside the bulk gap, as shown in Fig. 2(f)
and its inset. Thus, the 2D phononic SOTI is identified in the in-plane modes of graphdiyne.

%\begin{SCfigure*}
\begin{figure}
\includegraphics[width=8cm]{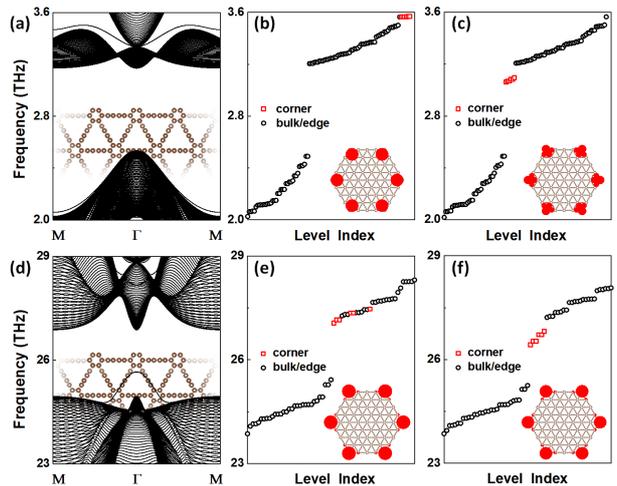}
\caption{(a) Out-of-plane phonon bands of graphdiyne nanoribbon with edge termination shown in the inset.
(b) Discrete out-of-plane phonon levels of hexagonal graphdiyne cluster with spatial distribution of
six corner states shown in the inset. (c) is the same to (b) but with an additional local corner potential that
can tune the energies of six corner states into bulk gap. (d)-(f) are the same to (a)-(c), but for
the in-plane modes of graphdiyne.}
\end{figure}
%\end{SCfigure}

The in-gap topological corner states provide a smoking gun signature
to identify the 2D phononic SOTI. However, the nontrivial bulk topology only guarantees the existence of
corner states, but doesn't restrict their energies \cite{45}, namely, the corner states may either inside
or outside the bulk gap. Additional symmetry is needed to pin them inside the bulk gap, such as chiral or
particle-hole symmetry \cite{45}. But these symmetries are not necessary to protect the 2D SOTI,
which may be broken in different edge configurations. To investigate the termination dependence of
topological corner states in graphdiyne, four more edges are considered, all showing the similar features (see Fig. S1 \cite{SI}).
For the edge terminated by bare benzene ring (inset of Fig. 3(a)), the out-of-plane phonon bands of
graphdiyne nanoribbon are shown in Fig. 3(a). Compared to Fig. 2(b), the edge state is pushed
upward to the bulk bands, while one lower edge state appears with slightly splitting from the bulk bands.
The corresponding discrete out-of-plane phonon levels of hexagonal graphdiyne cluster are shown in Fig. 3(b).
Compared to Fig. 2(c), the six corner states (red cube) are pushed upward to the bulk levels,
but they are still spatially localized at six corners (inset of Fig. 3(b)). Furthermore, if a small local
corner potential is added to the corner region, these localized corner states can be pushed back to the
bulk gap again, as shown in Fig. 3(c). Experimentally, this local perturbation method has already been used
to detect the topological corner states buried in the bulk states \cite{46}, making it easier to distinguish the
higher-order topological phase in different conditions. For the in-plane modes, the energies of edge and corner
states also depend on edge terminations, which are tunable by local corner potential, as shown in Fig. 3(d)-(f).
Hence, our results demonstrate the robust existence of phononic topological corner states in graphdiyne
cluster with different edge configurations, providing more convenience to realize them in the experiment.

%\begin{SCfigure*}
\begin{figure}
\includegraphics[width=8cm]{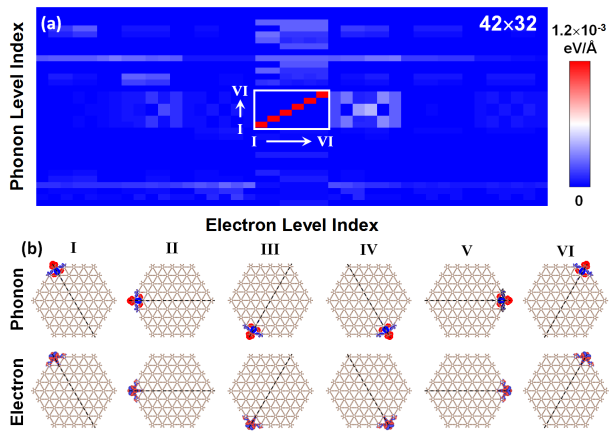}
\caption{(a) The EPC between 42 electron states around Fermi-level and 32 out-of-plane phonon modes around nontrivial gap.
The six phononic and electronic corner states are labeled as I to VI. The EPC between them is highlighted in the white rectangular region,
showing an enhanced EPC in six diagonal elements with the same index for phononic and electronic corner states.
(b) Spatial distribution of the localized six phononic and electronic corner states. The red/blue color denotes positive/negative value
of phonon mode and electron state. The black-dashed line denotes three in-plane mirror planes.}
\end{figure}
%\end{SCfigure}

Since the phononic and electronic topological corner states coexist in graphdiyne,
the EPC is further calculated to reveal the unique interplay
between topological phonon and electron. The EPC ($\Xi_\nu^n$) between one electron
state ($n$) and one phonon mode ($\nu$) is defined as
%\begin{equation}
$\Xi_\nu^n=\langle\psi_n|\Delta H_\nu|\psi_n\rangle$
%\end{equation}
\cite{47,SI}, where $|\psi_n\rangle$ is the electron eigenstate with index $n$ at equilibrium
configuration, and $\Delta H_\nu$ is the variation of electron Hamiltonian induced
by phonon mode $\nu$. Before performing the numerical calculations, we first make a
qualitative analysis about this EPC. Based on the definition of $\Delta H_\nu$,
one can see that its non-zero elements are limited in the space where the phonon mode
$\nu$ is non-zero. Similarly, the non-zero elements of $|\psi_n\rangle$ are also limited
in the space where the electron eigenstate $n$ is non-zero. Therefore, the degree of overlapping
between these two spaces will determine the intensity of EPC. This indicates that the
enhanced EPC may be achieved for phononic and electronic corner states localized at the
same corner. Moreover, there are three in-plane mirror planes in graphdiyne, so the phonon
mode has either even or odd mirror symmetry. Our two-atom model analysis shows that the EPC
will be canceled by the phonon mode with odd mirror symmetry, which is irrelevant to the
symmetry of electron state \cite{SI}. This indicates that the enhanced EPC may be achieved for
phononic corner states with even mirror symmetry only.

To check this analysis, the coupling between 42 electron states around
Fermi-level (see Fig. S2 \cite{SI}) and 32 out-of-plane phonon modes around nontrivial gap (Fig. 2(c))
are calculated, as shown in Fig. 4(a). The EPC between six phononic and electronic corner
states (labeled as I to VI) are highlighted in the while rectangular region, including 36 elements.
The spatial distribution of the phononic and electronic corner states are shown in Fig. 4(b),
exhibiting even mirror symmetry. Obviously, if the phonon and electron are localized at the same corner,
the EPC is large (red color), otherwise, it is small (blue color). Moreover, the EPC between electron
states and in-plane phonon modes (Fig. 2(f)) are also calculated. In this case, all phononic corner states
have odd mirror symmetry, and no enhanced EPC is observed for phonon and electron localized at the
same corner (see Fig. S3 \cite{SI}). These features are consistent with our qualitative analysis.
Consequently, a unique selectivity of space and symmetry in EPC is revealed
for the topological corner states.

In summary, we demonstrate the coexistence of 2D phononic and electronic SOTI in graphdiyne,
and discover a distinctive EPC between phononic and electronic topological corner states.
Our results extend the higher-order topological physics in Kekul\'{e} lattice and
introduce an exotic platform to explore the interplay between higher-order topological
phonon and electron, which will draw immediate experimental attentions.

This work was supported by National Natural Science Foundation of China (Grant No. 12174369, 11774325 and 21603210),
National Key Research and Development Program of China (Grant No. 2017YFA0204904),
and Fundamental Research Funds for the Central Universities.
We thank Supercomputing Center at USTC for providing the computing resources.

%\newpage{}%

\end{document}